\documentclass[twoside,12pt]{article}
\usepackage{epsfig}
\usepackage{epsfig,graphicx}
\usepackage{amsmath}
\usepackage{latexsym}

\newcommand{\be}{\begin{equation}}
\newcommand{\ee}{\end{equation}}
\newcommand{\bea}{\begin{eqnarray}}
\newcommand{\eea}{\end{eqnarray}}

\newcommand{\condensate}[1]{\ensuremath{\left \langle #1 \right \rangle}}
\newcommand{\vaccondensate}[1]{\ensuremath{\left \langle #1 \right \rangle_{\rm vac}}}
\newcommand{\borel}{\ensuremath{ \mathcal{M} }}
\renewcommand{\slash}[1]{\ensuremath{ #1 \mspace{-8mu} / }}
\newcommand{\vackappa}[1]{\ensuremath{\kappa^{\rm vac}_{ \rm #1 }}}
\newcommand{\medkappa}[1]{\ensuremath{\kappa^{\rm med}_{ \rm #1 }}}

\newcommand{\tmedkappa}[1]{\ensuremath{\tilde{\kappa}^{\rm med}_{ \rm #1 }}}
\newcommand{\tkappa}[1]{\ensuremath{\tilde{\kappa}}}

\begin{document}

\title{ \vspace{1cm} Role of Four-Quark Condensates in QCD Sum Rules}

\author{
R. Thomas$^1$, T. Hilger$^{2}$, B. K\"ampfer$^{1,2}$\\
\\
$^1$ Forschungszentrum Dresden-Rossendorf, PF 510119, 01314 Dresden, Germany\\
$^2$ Institut f\"ur Theoretische Physik, TU Dresden, 01062 Dresden, Germany}

\maketitle

\begin{abstract}
The QCD sum rule approach to the in-medium behavior of hadrons
is discussed for $\omega$ meson, nucleon and $D$ meson.
Emphasis is devoted to the impact of four-quark condensates
and to order parameters of spontaneous symmetry breaking.
\end{abstract}
%\pacs{12.40.Yx, 12.38.Lg\\
%Keywords: Hadrons(omega meson, nucleon, D meson), 
%Medium Modifications, QCD Sum Rules,
%Chiral Symmetry, Four-Quark Condensates}

\section{Introduction}\label{sec:intro}

QCD sum rules \cite{SVZ} represent one approach towards understanding the masses of hadrons.
Hadronic parameters, among them the masses, are linked to expectation values
of QCD operators, the condensates. The nonzero values of condensates point to a complicated
structure of the QCD vacuum, i.e., of the ground state of strong interaction.
There are a few particularly important condensates:
(i) The chiral condensate, $\langle \bar q q \rangle$,
is connected with the spontaneous breaking of chiral symmetry. Together with the explicit
chiral symmetry breaking by nonzero quark masses $m_q$, the pion mass $m_\pi$ is related
via the Gell--Mann-Oakes-Renner relation
$m_\pi^2 f_\pi^2 \propto - m_q \langle \bar q q \rangle$
\cite{Colangelo} to the pion decay constant $f_\pi$ and the chiral condensate.
(ii) The gluon condensate $\langle (\alpha_s / \pi) G_{\mu \nu} G^{\mu \nu} \rangle$
(with $G_{\mu \nu}$ as chromodynamic field strength tensor; 
$\alpha_s = g_s^2 /(4 \pi)$ is the strong coupling)
is anchored in the QCD trace anomaly and the breaking of the dilatation symmetry.
(iii) The quark-gluon condensate
$\langle \bar q g_s \sigma_{\mu \nu} G^{\mu \nu} q \rangle \propto \langle \bar q q \rangle$
may also be considered as order parameter
of chiral symmetry \cite{qg_cond}. There are (infinitely) many other condensates which may be
ordered according to their mass dimension. These quantities occur in the evaluation of the
current-current correlator by means of an operator product expansion (OPE).
Since organized as corrections in increasing powers of inverse hadron momentum, 
the (poorly known) higher-order condensates are expected to be of minor importance, 
so that the few low-dimension condensates suffice.

\begin{figure}[htb]
\begin{center}
%\begin{minipage}[t]{8 cm}
\vspace*{-0.1cm}
\includegraphics[width=6cm,angle=270]{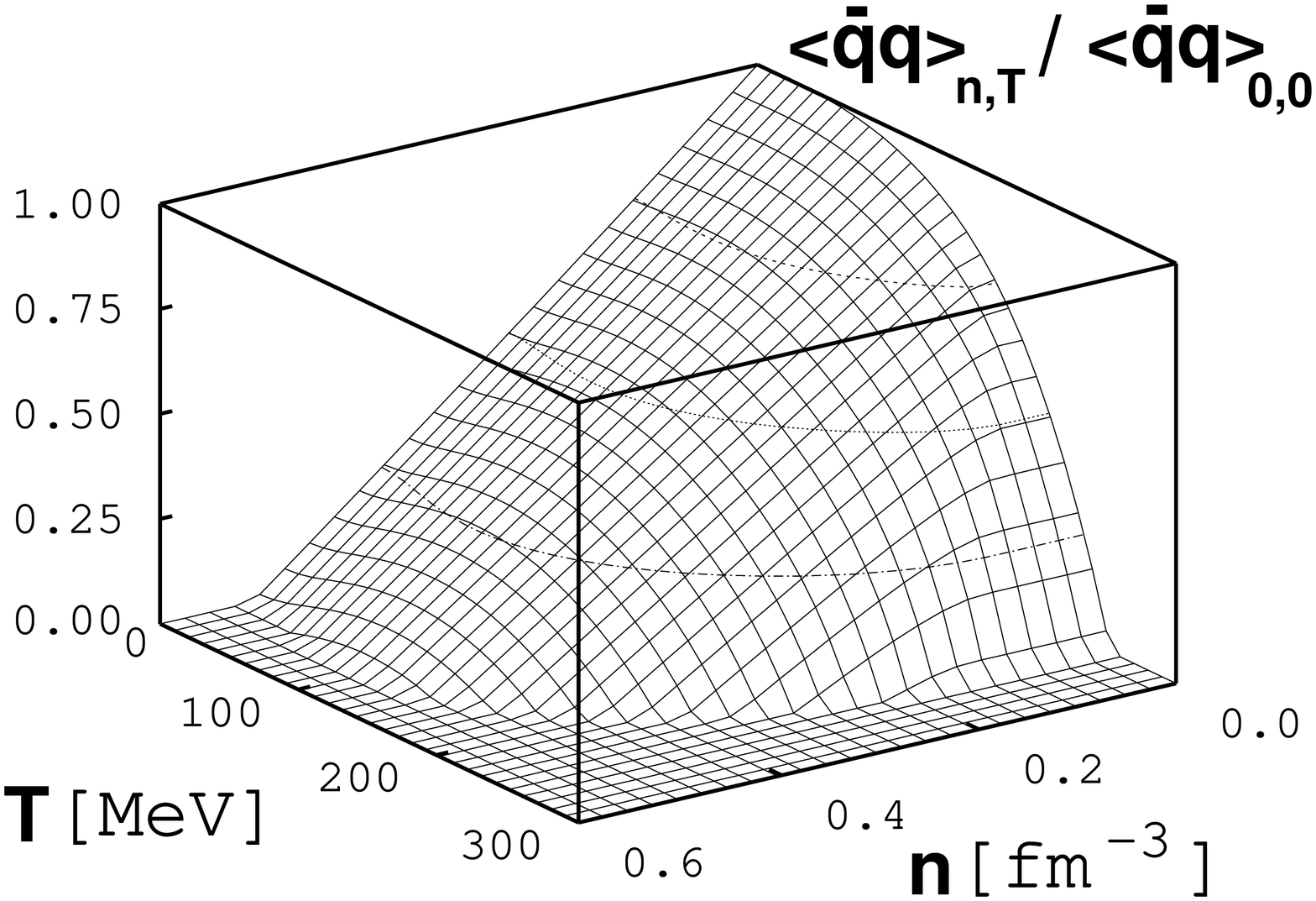}
\includegraphics[width=6cm,angle=270]{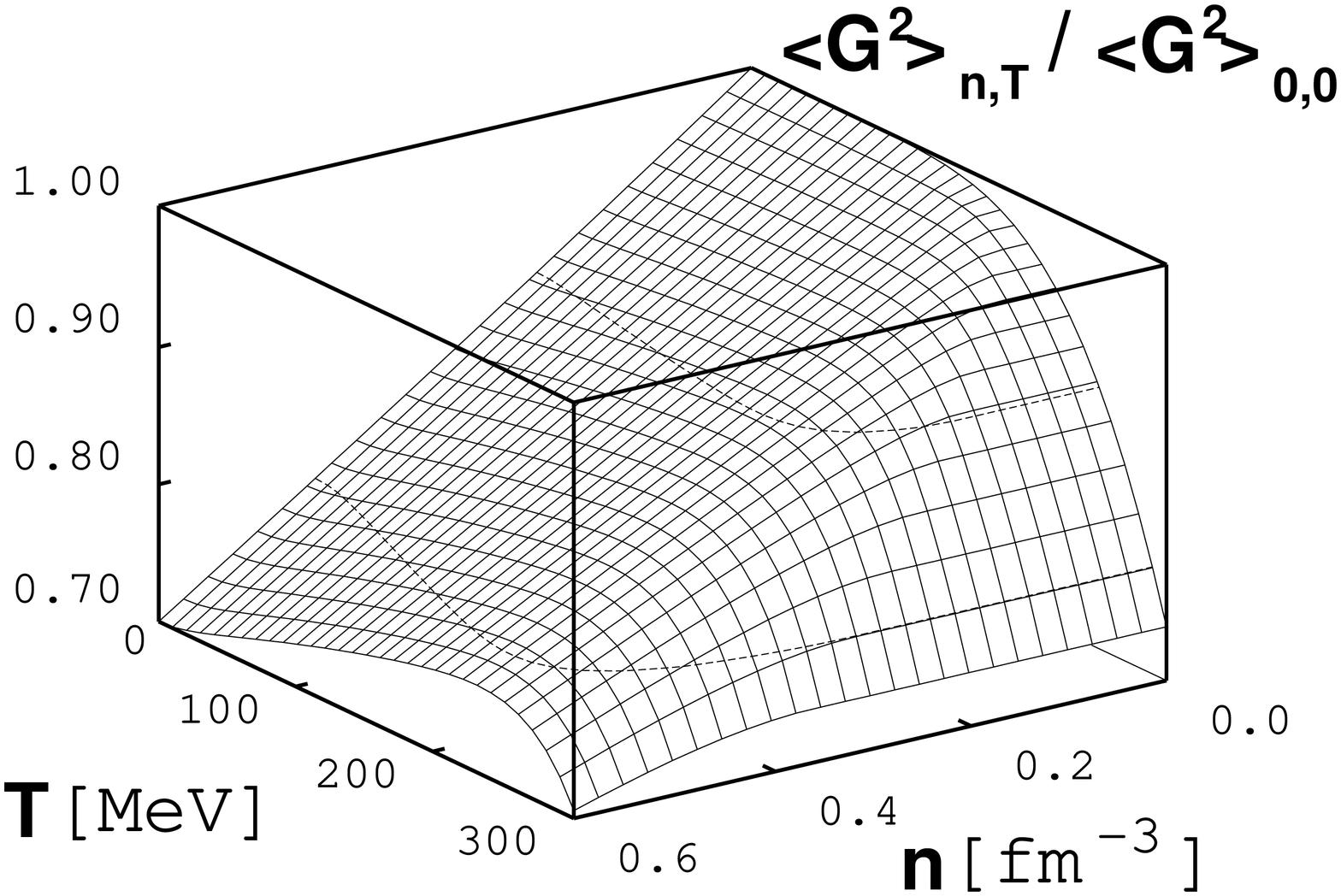}
\vspace*{-0.01cm}
%\end{minipage}
\begin{minipage}[t]{16.5 cm}
\caption{\label{fig.1}
Density and temperature dependence of chiral condensate (left panel) 
and gluon condensate (right panel). The employed approximation is reliable only for
small values of $T$ and $n$ \cite{Bormio}. Note the different slopes in
$n$ direction.}
\end{minipage}
\end{center}
\end{figure}

However, there are special circumstances, for instance in the 
Borel transformed QCD sum rules for
light vector mesons, in which the low-order condensates are numerically suppressed.
Specifically, the chiral condensate enters in the renormalization invariant
combination $m_q \langle \bar q q \rangle$ being a tiny contribution due to
the smallness of $m_q$. Instead, the four-quark condensates become important.
In a flavor-symmetric vacuum, there are 10
independent four-quark condensates \cite{NPA_2007}, all with highly unsettled
values. The situation becomes even more dramatic when extending the QCD sum rule
approach to a strongly interacting medium \cite{BS}. At finite baryon density $n$ and
temperature $T$ the condensates are modified, as exhibited in Fig.~\ref{fig.1} for the
chiral and the gluon condensate, 
and further condensate structures arise.
This modification is expected to be
reflected in hadron properties \cite{Hatsuda_Lee}. In particular, the
approach to chiral symmetry restoration was thought to be measurable by
investigating hadrons embedded in a strongly interacting medium \cite{BR}.
The QCD sum rule approach provides, however, a more involved relation.
The in-medium modifications of light vector mesons depend crucially on the
density dependence of a certain combination of four-quark condensates
\cite{PRL_2005}.
In a strongly interacting medium, there are 44
independent four-quark condensates in the $u, \, d$ sector \cite{NPA_2007}; 
if flavor symmetry is satisfied
this number reduces to 20. These enter the QCD sum rules
for different hadrons in different combinations. Therefore, one can hope that
a systematic investigation of various hadrons may shed light on the
important four-quark condensate combinations.

Formally, four-quark condensates are QCD ground state expectation values
of Hermitian products of four quark operators which are to be Dirac and Lorentz
scalars, color singlets and are to be invariant under parity and time reversal
(the latter for equilibrated systems).
Physically, the four-quark condensates quantify the correlated production
of two quark-antiquark pairs in the physical vacuum.
In contrast to the square of the chiral (two-quark) condensate,
which accounts for uncorrelated production of two of these pairs,
the four-quark condensates are a measure of the correlation
and thus expose the complexity of the QCD ground state.
Especially, deviations from factorization, i.e.\
the approximation of unknown four-quark condensates in
terms of the squared chiral condensate, justified in the large $N_c$ limit
\cite{Leupold:2005eq}, represent effects of these more involved correlations.

As mentioned above, the chiral condensate $\langle \bar q  q \rangle$
is often considered an order parameter for the $SU(N_f)_A$
chiral symmetry of QCD and, therefore, its density and temperature
dependence is investigated extensively.  
The change of $\langle \bar q  q \rangle$, however, might partially originate from
virtual low-momentum pions and thus could not clearly signal partial restoration
of chiral symmetry in matter \cite{Birse}.
The interpretation of four-quark condensates as order parameters
for spontaneous breaking of chiral symmetry is still an open issue.
In \cite{Leupold:2006ih} a specific combination of four-quark condensates
arising from the difference between vector and axial-vector correlators
is proposed as such an alternative parameter. This combination
(shown to agree with vacuum factorization in the analysis of
$\tau$ decay data \cite{Bordes:2005wv}) is distinct from the four-quark
condensates in the nucleon channel as well as from the combination in the
$\omega$ sum rule \cite{NPA_2007}.
For instance, in vacuum QCD sum rules for the nucleon \cite{NPA_2007}
the four-quark condensate combination contains a chirally invariant part
when assuming flavor symmetry ($\psi$ is a flavor vector)
\begin{equation}
\left [ 2(2t^2+t+2) \condensate{\bar{\psi} \gamma_\mu \psi \bar{\psi} \gamma^\mu \psi}
+ (3t^2+4t+3) \condensate{\bar{\psi} \gamma_5 \gamma_\mu \psi \bar{\psi}
\gamma_5 \gamma^\mu \psi} \right ] - \frac34
\left [\text{color structures with} \, \lambda^A \right], \nonumber
\end{equation}
since the condensates $\condensate{\bar{\psi} \gamma_\mu \psi \bar{\psi} \gamma^\mu \psi}$
and $\condensate{\bar{\psi} \gamma_5 \gamma_\mu \psi \bar{\psi}
\gamma_5 \gamma^\mu \psi}$ are invariants w.r.t.\ the $SU(N_f)_A$
transformation $\psi \to \psi' = \exp\{i \beta_\alpha T^\alpha \gamma_5 \} \psi$
and contains another part
\begin{equation}
\left[ 3(t^2-1) \left ( \condensate{\bar{\psi} \psi \bar{\psi} \psi} +
\condensate{\bar{\psi} \gamma_5 \psi \bar{\psi} \gamma_5 \psi} - \frac12
\condensate{\bar{\psi} \sigma_{\mu \nu} \psi \bar{\psi} \sigma^{\mu \nu} \psi} \right ) \right]
- \frac34 \left [\text{color structures with} \, \lambda^A \right] \nonumber
\end{equation}
including condensates which break the $SU(N_f)_A$ symmetry. 
The vanishing of the four-quark condensates in this second part would restore
the broken $SU(N_f)_A$ symmetry. 
The combinations are weighted by polynomials of the mixing parameter $t$
allowing general forms of the nucleon interpolating field.
In the preferred case $t=-1$ (the Ioffe current for the nucleon)
only the chirally invariant part survives
and thus this remainder cannot be used as an order parameter.
Additional insight into the change of four-quark condensates
and their role as order parameters of spontaneous chiral symmetry breaking
could be acquired from other hadronic channels.

Nevertheless, when identifying chiral symmetry restoration with
degenerating vector ($V$) and axial-vector ($A$) spectral functions
the four-quark condensates occur as order parameter~\cite{Hatsuda_Lee,Kapusta_Shuryak}
\begin{equation}
\int dq \, q^3 (\Pi_V(q) - \Pi_A(q)) = - \frac12 \pi \alpha_s \langle {\cal O}_4 \rangle
\end{equation}
with
\begin{eqnarray}
\langle {\cal O}_4 \rangle &=& 
\langle (\bar u \gamma_\mu \gamma_5 \lambda^a u 
- \bar d \gamma_\mu \gamma_5 \lambda^a d)^2 \rangle
-
\langle (\bar u \gamma_\mu \lambda^a u 
- \bar d \gamma_\mu \lambda^a d)^2 \rangle \nonumber
\end{eqnarray}
showing (i) that four-quark condensates are important for chiral symmetry
restoration in the sense of degeneracy of vector and axial-vector currents,
and (ii) that such formulations refer to moments, i.e., weighted integrals.
 
Our contribution is organized as follows. In section~\ref{sec:disprel}, 
we consider a dispersion
relation being the starting point of the QCD sum rule approach.
Section~\ref{sec:probes} is devoted to specific cases: The status of the 
$\omega$ meson is recapitulated~(\ref{sec:probes}.1) 
and the nucleon is discussed~(\ref{sec:probes}.2). 
The case of the $D$ meson as an example for the
light-heavy flavor sector is dealt with~(\ref{sec:probes}.3, 
where a brief remark is made 
on the $J/\psi$ as example for the heavy-heavy sector).
The summary can be found in section~\ref{sec:summary}.

\section{Contemplation on Dispersion Relations}\label{sec:disprel}

Let be given the analytical properties of the correlation function $\Pi (q_0)$ 
of a selected hadron (actually, all hadronic excitations with a given set
of the same quantum numbers),
such that it can be expressed through its spectral function 
$\Delta \Pi (\omega)$ via the dispersion relation
\begin{equation}
\label{eq:dispersionRelationAnsatz}
\frac{1}{\pi} \int_{-\infty}^{+\infty} d\omega
\frac{\Delta \Pi (\omega) }{\omega - q_0} = \Pi (q_0) .
\end{equation}
This leads to the celebrated form employed in QCD sum rules. The l.h.s.\ contains the
hadronic spectral function, while the r.h.s.\ is evaluated, for large Euclidian
energies $q_0$ and fixed momenta $\vec q$,
by means of the OPE. We assume for the moment being that the r.h.s.\
is settled. In order to attempt the isolation of the contribution of the
lowest hadronic excitation one can perform the following chain of reformulations.
Define even (''e'') and odd (''o'') contributions
\begin{eqnarray}
\dfrac{1}{\pi} \int_{-\infty}^{+\infty} d\omega
\dfrac{{\omega \Delta} \Pi (\omega) }{\omega^2 - q_0^2} &=&
\Pi^{\mathrm e} (q_0^2) \equiv \dfrac{1}{2} \left ( \Pi (q_0) + \Pi (-q_0) \right) ,\\
\dfrac{1}{\pi} \int_{-\infty}^{+\infty} d\omega
\dfrac{\Delta \Pi (\omega) }{\omega^2 - q_0^2} &=&
\Pi^{\mathrm o} (q_0^2) \equiv \dfrac{1}{2q_0}
\left ( \Pi (q_0) - \Pi (-q_0) \right )
\end{eqnarray}
and split the integrals with respect to the low-lying hadron
under consideration in the
intervals $0 \cdots \omega_+$ for positive energy and $\omega_- \cdots 0$ for
negative energy
\begin{eqnarray}
&&\dfrac{1}{\pi}\int_{\omega_-}^0 d \omega \omega \dfrac{\Delta \Pi (\omega)}{\omega^2 - q_0^2} +
\dfrac{1}{\pi}\int_0^{\omega_+} d \omega \omega \dfrac{\Delta \Pi (\omega)}{\omega^2 - q_0^2} \\
&& \hspace*{3cm} = \Pi^{\mathrm e}(q_0^2) -
\dfrac{1}{\pi}\int_{- \infty}^{\omega_-} d \omega \omega 
\dfrac{\Delta \Pi (\omega)}{\omega^2 - q_0^2} -
\dfrac{1}{\pi}\int_{\omega_+}^\infty d \omega \omega 
\dfrac{\Delta \Pi (\omega)}{\omega^2 - q_0^2}
\equiv R_{\mathrm e} , \nonumber \\
&&\dfrac{1}{\pi}\int_{\omega_-}^0 d \omega \dfrac{\Delta \Pi (\omega)}{\omega^2 - q_0^2} +
\dfrac{1}{\pi}\int_0^{\omega_+} d \omega \dfrac{\Delta \Pi (\omega)}{\omega^2 - q_0^2} \\
&& \hspace*{3cm} = \Pi^{\mathrm o}(q_0^2) -
\dfrac{1}{\pi}\int_{- \infty}^{\omega_-} d \omega \dfrac{\Delta \Pi (\omega)}{\omega^2 - q_0^2} 
-
\dfrac{1}{\pi}\int_{\omega_+}^\infty d \omega \dfrac{\Delta \Pi (\omega)}{\omega^2 - q_0^2}
\equiv R_{\mathrm o} . \nonumber
\end{eqnarray}
The integrals on the r.h.s.\ can be converted into expressions
similar to the OPE by exploiting the semi-local duality hypothesis.
Defining further the moments
\begin{equation}
\label{eq:ebardefinition}
E_- = \frac{\int_{\omega_-}^0 d\omega \omega \Delta \Pi (\omega) (\omega^2 - q_0^2)^{-1}}
{\int_{\omega_-}^0 d\omega \Delta \Pi (\omega) (\omega^2 - q_0^2)^{-1}}
\quad \text{and} \quad
E_+ = \frac{\int^{\omega_+}_0 d\omega \omega \Delta \Pi (\omega) (\omega^2 - q_0^2)^{-1}}
{\int^{\omega_+}_0 d\omega \Delta \Pi (\omega) (\omega^2 - q_0^2)^{-1}},
\end{equation}
the even and odd sum rules can be rephrased to formulate relations which rather depend on
ratios than purely on absolute spectral integrals.
In a combined resulting sum rule
\begin{equation}
\dfrac{1}{\pi} \int_0^{\omega_+} d \omega \dfrac{\Delta \Pi (\omega)}{\omega^2 - q_0^2}
=
\dfrac{R_{\mathrm e} - E_- R_{\mathrm o}}{E_+ - E_-}
\end{equation}
the integral over the negative energy contribution seems to be ''eliminated''.
(This contribution, however, is still included in the ratio $E_-$.)

In the special case of an anti-symmetric spectral function,
$\Delta \Pi (\omega) = - \Delta \Pi (-\omega)$,
e.g. for $\rho^0$ and $\omega$, 
one obtains with $\omega_- = - \omega_+$
that $E_+ = - E_-$, $R_{\mathrm o} = 0$ and (using $s=\omega^2 $) 
$\int_0^{s_+} d s \tfrac{\Delta \Pi (s)}{s - q_0^2} = R_{\mathrm e}$.
Therefrom one can define a normalized moment for the Borel transform
\begin{equation}
\bar m^2 (n,{\cal M}^2,s_+) \equiv \frac{\int_0^{s_+} ds \; \Delta \Pi (s,n) \;
e^{-s/{\cal M}^2}}{\int_0^{s_+} ds \; \Delta \Pi (s,n) \; s^{-1} e^{-s/{\cal M}^2}},
\label{mass_parameter}
\end{equation}
where ${\cal M}$ is the Borel mass.
Determined by the ratio of $R_{\mathrm e}$ and its logarithmic derivative, 
this quantity is model independent
(but suffers from the ad hoc introduced continuum threshold $s_+$).
Its meaning becomes obvious for a pole ansatz, $\Delta \Pi (s) = F \delta(m^2 - s)$, where
$\bar m (n,{\cal M}^2,s_+) = m$ follows.

In the generic case, however, the excitation spectrum of particles and anti-particles
in a medium is asymmetric, and the dispersion relation (\ref{eq:dispersionRelationAnsatz})
alone seems not to allow separate model-independent statements about the low-lying
excitation at positive energy. This appears as a severe limitation of this form of the
QCD sum rule approach.

\section{Distinct Probes for QCD Condensates}\label{sec:probes} 

\subsection{{\it {\bf $\omega$} meson}}\label{subsec:omega}

\begin{figure}[h]
\begin{center}
\begin{minipage}[t]{16.5 cm}
\begin{minipage}[t]{6cm}
\includegraphics[width=6cm,angle=-90]{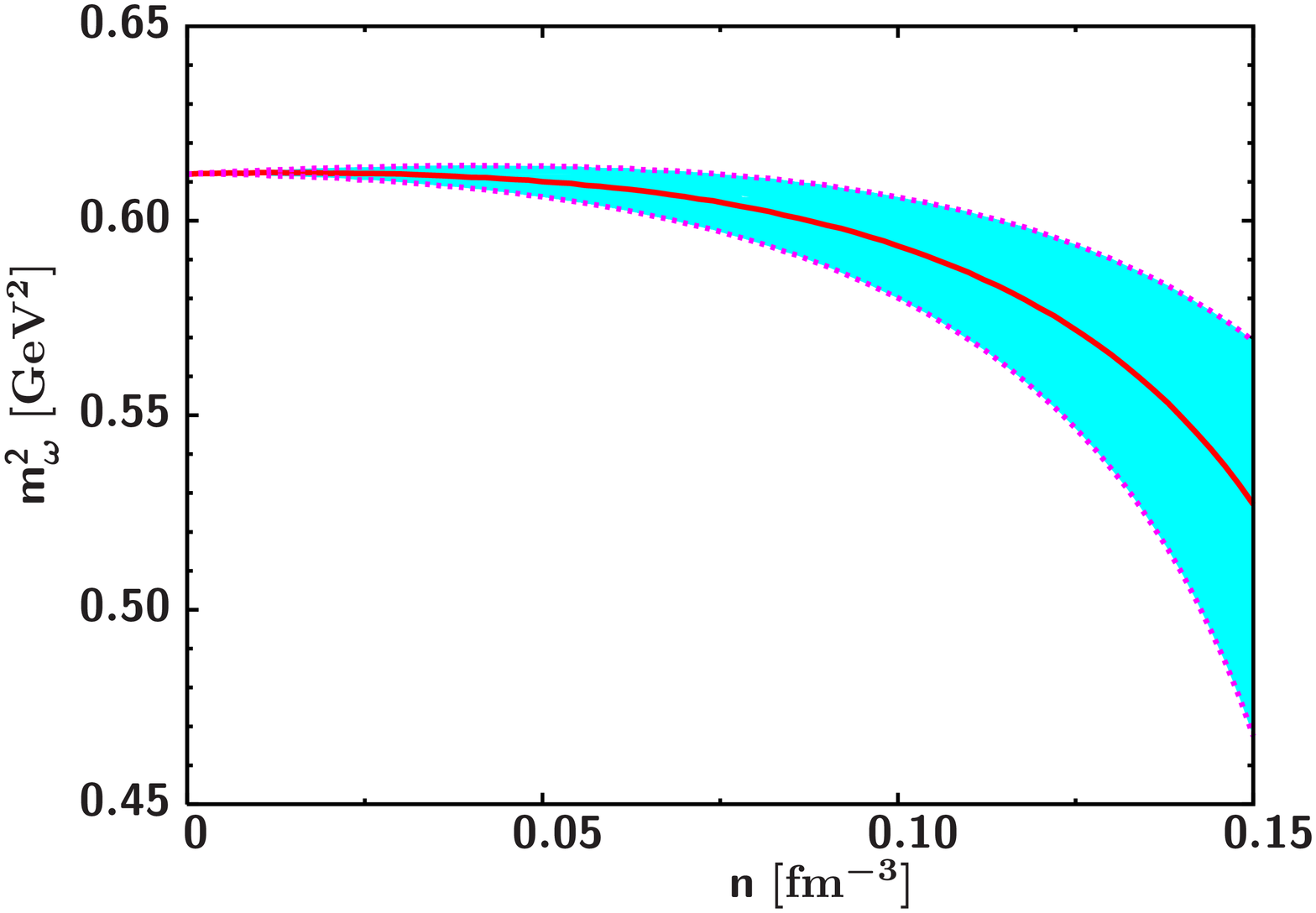}
\end{minipage}
\hfill
\begin{minipage}[t]{7.3cm}
\vskip -3mm
\caption{\label{fig.2}
The mass parameter defined in (\ref{mass_parameter})
and averaged within the Borel window as a function of the baryon density $n$ at $T = 0$
for $\kappa_\omega^{med} = 4$ and condensates up to mass dimension 6 (solid curve).
The effect of mass dimension 8 condensates
is exhibited by the shaded area.
For the sigma term $\sigma_{\rm N}$ and the vacuum condensate 
$\vaccondensate{ \bar{q} q} \equiv \langle \bar q q \rangle_{0,0}$
we employ standard values, while $\vackappa{\omega}$ is adjusted to the vacuum
$\omega$ mass.
For details cf.\ \cite{PRL_2005}.}
\end{minipage}
\end{minipage}
\end{center}
\end{figure}

The $\omega$ and $\rho^0$ mesons, including their mixing,
have been dealt with in \cite{PRL_2005,SZ_2004} in detail.
In \cite{PRL_2005} the combined four-quark condensates
contributing to the OPE terms %in the Borel mass power ${\cal M}^6$,
\begin{align}
&\dfrac{1}{2} \condensate{ \bar{u} \gamma_5 
\gamma_\mu \lambda^A u \bar{u} \gamma_5 \gamma^\mu \lambda^A u } 
+ \dfrac{1}{2} \condensate{ \bar{d} \gamma_5 \gamma_\mu \lambda^A d \bar{d} 
\gamma_5 \gamma^\mu \lambda^A d }
+ \condensate{ \bar{u} \gamma_5 \gamma_\mu \lambda^A u \bar{d} 
\gamma_5 \gamma^\mu \lambda^A d } \nonumber \\
+ & \dfrac{2}{9} \condensate{ \bar{u} \gamma_\mu  
\lambda^A u \bar{d} \gamma^\mu \lambda^A d } 
+ \dfrac{1}{9}  \condensate{ \bar{u} \gamma_\mu 
\lambda^A u \bar{u} \gamma^\mu \lambda^A u } + 
\dfrac{1}{9} \condensate{ \bar{d} \gamma_\mu \lambda^A d \bar{d} \gamma^\mu \lambda^A d } 
\nonumber \\ 
& \qquad = \dfrac{112}{81} \left(
\vackappa{\omega} \vaccondensate{ \bar{q} q }^2 + 
\medkappa{\omega} \vaccondensate{ \bar{q} q} \dfrac{\sigma_{\rm N}}{ m_{\rm q}} n \right)
\end{align}
in the QCD sum rule for the $\omega$ meson
have been assigned a strong density dependence, 
at least $\medkappa{\omega} = 4$.
Otherwise the numerically
large Landau damping term would push up the weighted strength \cite{Oleg}.
In contrast,
the experiment \cite{CB_TAPS} finds a lowering of the $\omega$ strength.
Fig.~\ref{fig.2} exhibits the density dependence of the mass parameter 
(\ref{mass_parameter}).
Note that this parameter coincides only in zero-width approximation
with the $\omega$ pole mass squared; in general it is a normalized
moment of ${\rm Im}\Pi^\omega$ to be calculated from data or models.

\subsection{{\it Nucleon}}\label{subsec:nucleon}

\begin{figure}[h]
\begin{center}
\begin{minipage}[t]{16.5 cm}
\begin{minipage}[t]{6cm}
\includegraphics[width=7cm,angle=-90]{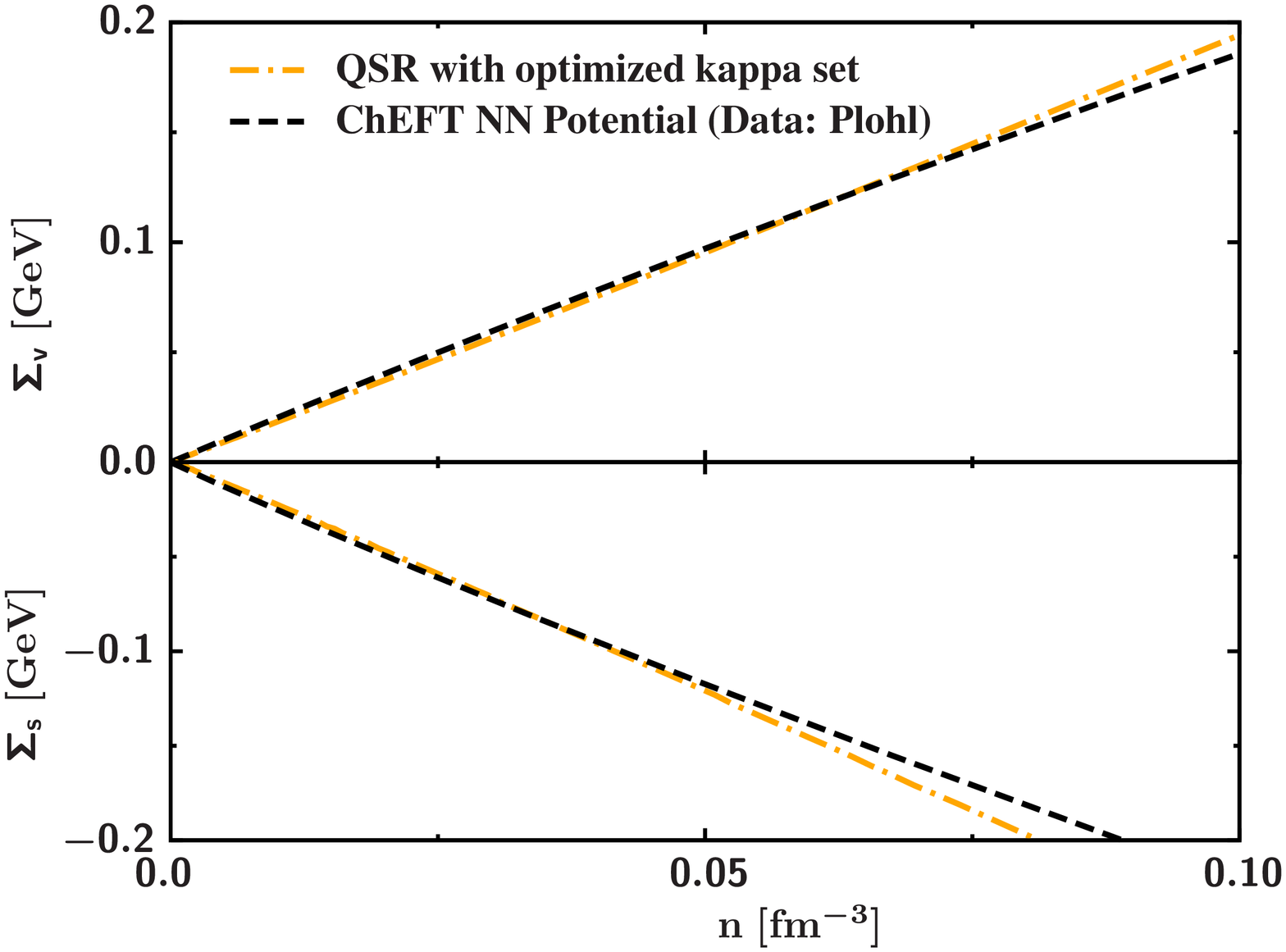}
\end{minipage}
\hfill
\begin{minipage}[t]{6cm}
\caption{\label{fig.3}
Nucleon vector and scalar self-energies as a function
of the baryon density $n$ at $T = 0$ for a special choice of 
$\medkappa{s}$, $\medkappa{q}$, $\tmedkappa{v}$, see \cite{NPA_2007}.
The self-energies from chiral effective field theory \cite{Plohl:2006hy}
(ChEFT, dash-dotted curves) are shown as well.}
\end{minipage}
\end{minipage}
\end{center}
\end{figure}

The nucleon QCD sum rule is more involved since several Dirac structures need
to be considered \cite{Furnstahl}:
$
\Pi (q) = \Pi_s (q^2,qv) + \Pi_q (q^2,qv) \slash{q} + \Pi_v (q^2,qv) \slash{v} ,
$
i.e., besides the even and odd parts one has
$\Delta \Pi_i$ with $i = s, q, v$,
where $v$ refers to the four-velocity of the medium. 
We evaluate the QCD sum rule by utilizing the ansatz
\begin{equation}
\Delta G (q_0) = \dfrac{\pi}{1-\Sigma_q} \dfrac{\slash{q} + M_N^* -
\slash{v} \Sigma_v}{E_+ - E_-} \left[ \delta (q_0 - E_- ) - \delta (q_0 - E_+ ) \right] ,
\end{equation}
with the implication
\begin{equation}
(E_+ - E_-) \frac{1}{\pi} \int_{0}^{\omega_+} 
d \omega \Delta \Pi (\omega) e^{-\omega^2 / 
\borel^2} =
- \frac{\lambda_N^2}{1-\Sigma_q} ( \slash{q} + M_N^* - 
\slash{v} \Sigma_v) e^{-E_+^2 / \borel^2} .
\end{equation}
The quantities $E_\pm = \Sigma_v \pm \sqrt{M_N^{* 2} + \vec q^{\,2}}$ are related, in the ansatz,
to the scalar and
vector contributions to the self-energy, $\Sigma_s = M_N^* - M_N$ and $\Sigma_v$.
Figure~\ref{fig.3} exhibits $\Sigma_{s,v}$
for special values of the density dependence parameters
$\kappa_s^{med} = 1.2$
$\kappa_q^{med} = -0.4$
$\tilde\kappa_v^{med} = 0.1$
entering the three independent four-quark condensate combinations
\begin{equation}
\condensate{ \bar{u} \slash{v} u \bar{d} d } + 
\dfrac{1}{2} \condensate{ \bar{u} \gamma_5 \gamma_\kappa u \bar{d} 
\sigma_{\lambda \pi} d \epsilon^{\kappa \lambda \pi \xi} v_\xi } 
- \dfrac{3}{4} \left [\text{color structures with} \, \lambda^A \right ]
=
\medkappa{s} \vaccondensate{ \bar{q} q} \dfrac{3}{2} n ,
\end{equation}
\vskip -6mm
\begin{align}
&
\condensate{ \bar{u} \gamma_\tau u \bar{u} \gamma^\tau u } - 
\condensate{ \bar{u} \slash{v} u \bar{u} \slash{v} u / v^2 }
- \condensate{ \bar{u} \gamma_5 \gamma_\tau u \bar{u} \gamma_5 
\gamma^\tau u } + \condensate{ \bar{u} \gamma_5 \slash{v} u \bar{u} \gamma_5 \slash{v} u / v^2 }
\nonumber \\
& + 
4 \condensate{ \bar{u} \gamma_\tau u \bar{d} \gamma^\tau d } - 
\condensate{ \bar{u} \slash{v} u \bar{d} \slash{v} d / v^2 }
+ 2 \condensate{ \bar{u} \gamma_5 \gamma_\tau u \bar{d} \gamma_5 \gamma^\tau d } + 
\condensate{ \bar{u} \gamma_5 \slash{v} u \bar{d} \gamma_5 \slash{v} d / v^2 }
 \nonumber \\
& - \dfrac34 \left [\text{color structures with} \, \lambda^A \right] 
= 
\dfrac32 \left(
\vackappa{q} \vaccondensate{ \bar{q} q }^2 + \medkappa{q} 
\vaccondensate{ \bar{q} q} \dfrac{\sigma_{\rm N}}{ m_{\rm q}} n \right),
\end{align}
\vskip -6mm
\begin{eqnarray}
&&
- \dfrac{1}{4} \condensate{ \bar{u} \gamma_\tau u \bar{u} \gamma^\tau u }
+ \condensate{ \bar{u} \slash{v} u \bar{u} \slash{v} u / v^2 } 
+ \dfrac{1}{4} \condensate{ \bar{u} \gamma_5 \gamma_\tau u \bar{u} \gamma_5 \gamma^\tau u }
- \condensate{ \bar{u} \gamma_5 \slash{v} u \bar{u} \gamma_5 \slash{v} u / v^2 }
\nonumber \\
&&- \dfrac{1}{4} \condensate{ \bar{u} \gamma_\tau u \bar{d} \gamma^\tau d } 
+ \condensate{ \bar{u} \slash{v} u \bar{d} \slash{v} d / v^2 }
+ \dfrac{1}{4} \condensate{ \bar{u} \gamma_5 \gamma_\tau u \bar{d} 
\gamma_5 \gamma^\tau d }
- \condensate{ \bar{u} \gamma_5 \slash{v} u \bar{d} 
\gamma_5 \slash{v} d / v^2 }  \nonumber \\
&& - \dfrac34 \left [\text{color structures with} \, \lambda^A \right ] 
=  
\dfrac32 
\tmedkappa{v} \vaccondensate{ \bar{q} q} \dfrac{\sigma_{\rm N}}{ m_{\rm q}} n .
\end{eqnarray}
We use further $\kappa_q^{vac} = 0.8$ and assume
equal continuum thresholds for the three components of the sum rule
for contributions $i = s, q, v$. % in (\ref{eq:invariantDecomposition}).
Changes of the self-energies under individual variations of 
$\medkappa{s}$, $\medkappa{q}$, $\tmedkappa{v}$ 
are reported in \cite{NPA_2007}. The variation of the genuine chiral condensate
$\langle \bar q q \rangle$ causes only tiny changes.

\subsection{{\it $D$ meson}}\label{subsec:dmeson}

The $D$ meson QCD sum rule is entered \cite{Hayashigaki,Morath} by the combination
$m_c \langle \bar q q \rangle$, i.e., the charm mass $m_c$ acts as magnifier 
of the chiral condensate. The evaluation of the sum rule
for the $D^\pm$, $D_0$, $\bar D_0$ multiplet involves the
renormalization of non-perturbative condensates in order to compensate the parts 
of perturbative
contributions, which are dominated by the non-vanishing mass scales. 
These parts are of non-perturbative
origin and thus should be covered by the appropriately defined non-perturbative condensates 
(usually called ''non-normal ordered'' condensates \cite{pk:Narison2004})
\begin{equation}
\langle \bar{q} \, \hat{O} \left( \partial_{\mu} - i g A_{\mu} \right) q \rangle 
=
\langle : \bar{q} \, \hat{O} \left( \partial_{\mu} - i g A_{\mu} \right) q : \rangle
-i\int d^4p  \left \langle Tr \left[
\hat{O} \left( -ip_{\mu} - i\tilde{A}_{\mu} \right) S^{\rm per}(p)
\right] \right \rangle,
\end{equation}
which quantify the non-perturbative regime \cite{pk:SpiridonovChetyrkin1988}. 
The integral denotes the perturbatively calculated contribution \cite{Grozin:1994hd}. 
The operator $\hat{O}$ is a function of the covariant derivative and contains general Dirac structures, 
$A$ is the gluon gauge field in fixed-point gauge and $\tilde{A}$ its Fourier transform; 
$: \cdots :$ denotes normal ordering, and $S^{\rm per}$ is the quark propagator 
in the gluonic background field.

Application of this condensate renormalization cancels mass logarithms in OPE calculations
\cite{Jamin:1992se} and causes a mixing between different types of condensates, 
e.g., the chiral condensate mixes with the gluon condensate (cf.\ also \cite{pk:Narison2004})
\begin{equation} \label{eq.17}
\begin{split}
    \langle\bar{q}q\rangle &= \langle:\bar{q}q:\rangle
         + \frac{3}{4\pi^2}m_q^3 \left( \ln{\frac{\mu^2}{m_q^2}}+1 \right)
         - \frac{1}{12m_q} \left \langle:\frac{\alpha_s}{\pi}G^2: \right \rangle + ...\quad,
\end{split}
\end{equation}
which relates in the heavy quark sector the heavy quark condensate 
(e.g.\ $\langle \bar c c \rangle$ for the charm condensate)
to gluon condensates
\cite{Grozin:1994hd,Generalis_Broadhurst}
\begin{equation}
m_c \langle \bar c c \rangle =
- \frac{1}{12} \left \langle \dfrac{\alpha_s}{\pi} G^2 \right \rangle -
\frac{1}{m_c^2}\frac{1}{1440 \pi^2} \langle g_s^3 G^3 \rangle + ...\quad .
\end{equation}
This is utilized in the QCD sum rule for the $J/\psi$ meson entered by the combination
$m_c \langle \bar c c \rangle$, which one expects to exhibit 
the weak density dependence of the gluon condensate, as shown in Fig.~1. In fact, 
the evaluation of the QCD sum rule for $J/\psi$ shows only a tiny change
of the in-medium mass \cite{Weise}.

In finite density QCD sum rules for the $D$ meson, medium-specific condensates 
appear and introduce, via their mixing, further gluon condensates into the OPE. 
An example is
\begin{equation}
\begin{split}
    \langle\bar{q}\gamma_{\mu}iD_{\nu}q\rangle &= \langle:\bar{q}\gamma_{\mu}iD_{\nu}q:\rangle
         + \frac{9}{4\pi^2}m_q^4 g_{\mu\nu} \left( \ln{\frac{\mu^2}{m_q^2}}
         + \frac{5}{12} \right) - \frac{g_{\mu\nu}}{48} \left \langle:\frac{\alpha_s}{\pi}G^2:\right \rangle
        \\ &
         + \frac{1}{18} \left( g_{\mu\nu} - 4\frac{v_\mu v_\nu}{v^2} \right)
         \left( \ln{\frac{\mu^2}{m_q^2}} - \frac{1}{3} \right)
         \left \langle:\frac{\alpha_s}{\pi} \left ( \frac{(vG)^2}{v^2} - \frac{G^2}{4} \right ) :\right \rangle,
\end{split}
\end{equation}
where $\mu$ is the renormalization scale, already introduced in (\ref{eq.17}).
The complete reevaluation of the in-medium 
$D$ meson QCD sum rules, in particular the complete OPE side up to mass dimension 6 
for products of quark masses and condensates, is under investigation.

\section{Summary}\label{sec:summary}

In summary we survey the QCD sum rules for the $\omega$ meson and the
nucleon. The sum rules allow for a direct relation of hadron properties
to QCD condensates which change at nonzero baryon density.
The genuine chiral condensate has numerically a small influence in the
Borel transformed sum rules. Instead, the four-quark condensates determine
to a large extent the density dependence of these hadrons composed of light
quarks. In contrast, in the heavy-light quark sector, e.g., represented by $D$ mesons,
the impact of the chiral condensate is expected to become noticeable.
A systematic investigation of various hadron species is needed to 
investigate the role of
the many four-quark condensates entering the QCD sum rules in
different combinations.

\subsubsection*{Acknowledgements}
The work is supported by BMBF 06DR136, GSI-FE and EU I3HP.

\end{document}